\documentclass[%
 aip,
rsi,%
 amsmath,amssymb,
 reprint,%
]{revtex4-1}

\usepackage{graphicx}
\usepackage{dcolumn}
\usepackage{bm}

\begin{document}

\title{Two-color optically-addressed spatial light modulator as generic spatio-temporal systems}

\author{V. Semenov}
\email{vladimir.semenov@femto-st.fr}
\affiliation{FEMTO-ST Institute/Optics Department, CNRS \& University Bourgogne Franche-Comt\'e, \\15B avenue des Montboucons,
	Besan\c con Cedex, 25030, France}
\author{X. Porte}%
\affiliation{FEMTO-ST Institute/Optics Department, CNRS \& University Bourgogne Franche-Comt\'e, 
	\\15B avenue des Montboucons,
	Besan\c con Cedex, 25030, France
}%
\author{I. Abdulhalim}
\affiliation{Department of Electro-Optics and Photonics Engineering, ECE School, and the Ilse Katz Institute for Nanoscale Science and Technology,\\
	Ben-Gurion University of the Negev, 84105 Beer-Sheva, Israel
}%
\author{L. Larger}
\affiliation{FEMTO-ST Institute/Optics Department, CNRS \& University Bourgogne Franche-Comt\'e, \\15B avenue des Montboucons,
	Besan\c con Cedex, 25030, France
}%
\author{D. Brunner}
\affiliation{FEMTO-ST Institute/Optics Department, CNRS \& University Bourgogne Franche-Comt\'e, \\15B avenue des Montboucons,
	Besan\c con Cedex, 25030, France
}%

\date{\today}

\begin{abstract}

Nonlinear spatio-temporal systems are the basis for countless physical phenomena in such diverse fields as ecology, optics, electronics and neuroscience.
The canonical approach to unify models originating from different fields is the normal form description, which determines the generic dynamical aspects and different bifurcation scenarios.
Realizing different types of dynamical systems via one experimental platform that enables continuous transition between normal forms through tuning accessible system parameters is therefore highly relevant.
Here, we show that a transmissive, optically-addressed spatial light modulator under coherent optical illumination and optical feedback coupling allows tuning between pitchfork, transcritical and saddle-node bifurcations of steady states.
We demonstrate this by analytically deriving the system's normal form equations and confirm these results via extensive numerical simulations.
Our model describes a nematic liquid crystal device using nano-dimensional dichalcogenide (a-As$_2$S$_3$) glassy thin-films as photo sensors and alignment layers, and we use device parameters obtained from experimental characterization.
Optical coupling, for example using diffraction, holography or integrated unitary maps allow implementing a variety of system topologies of technological relevance for neural networks and potentially XY-Hamiltonian models with ultra low energy consumption.

\end{abstract}

\maketitle

\begin{quotation}
	A broad variety of single or coupled dynamical systems can be experimentally realized by means of spatial light modulators.
	Optically-addressed spatial light modulators (OASLMs) inherently react to optical input signals and therefore allow for autonomous dynamical systems, which simplifies construction and reduces energy consumption.
	Furthermore, they provide straight-forward implementation of optical networks exploiting coherent optical feedback effects.
	This is attractive for parallel, scalable and energy efficiency neural networks.
	Furthermore, here we show that coherent feedback allows to implement a wide range of nonlinear oscillator networks exhibiting highly diverse yet tunable bifurcation scenarios.
	We analytically derive the conditions for implementing pitchfork, transcritical and saddle-node bifurcations of steady states without modifying the system under study and simply by tuning the relative optical intensities of our two-color illumination.
\end{quotation}

\section{Introduction}
\label{intro}
Dynamics exhibited by photonic nonlinear oscillator systems are very diverse and include regular and chaotic self-oscillatory behaviour \cite{bowden1984}, stochastic resonance \cite{gammaitoni1998}, coherence resonance \cite{giacomelli2000,koester2021}, noise-induced transitions \cite{naseri2015,barberoshie1993} or complex spatial structures revealed in the temporal dynamics of delay-feedback oscillators \cite{larger2013,giacomelli2012,brunner2018}.
An inherent asset of such optical systems is its high bandwidth, which makes them attractive for practical applications such as optical communication \cite{lavrov2010,argyris2005} and signal processing \cite{willner2014}.
However, fundamental appeal of photonic architectures is their almost unlimited parallelism \cite{Psaltis1990} combined with their unique potential for information transduction \cite{Lohmann1990}.
All these features make photonic architectures prime candidates for novel implementation and technological exploitation \cite{Genty2021} of large scale systems, and for network-based concepts in particular \cite{bueno2018,Wetzstein2020,Zhou2021,Porte2021}.

One compelling strategy for large-scale optical system synthesis are spatial light modulators (SLMs).
The excellent scalability of SLMs (commercial devices now enable up to $10^7$ individual oscillators) make them suitable for creating spatially-extended systems with large-scale ensembles of interacting oscillators \cite{residori2005}.
Besides the importance associated with observing different families of complex dynamics in physical experiments, such systems are of practical relevance for applications, for example in novel information processing concepts \cite{Wetzstein2020}.
Consequently, SLMs have been widely applied in the context of machine learning for the photonic neural network development \cite{bueno2018,rafayelyan2020,Zhou2021} as well as for the solution of combinatorial problems by the implementation of photonic Ising machines \cite{pierangeli2019,pierangeli2020}. 

\begin{figure*}[t]
	\centering
	\includegraphics[width=0.9\textwidth]{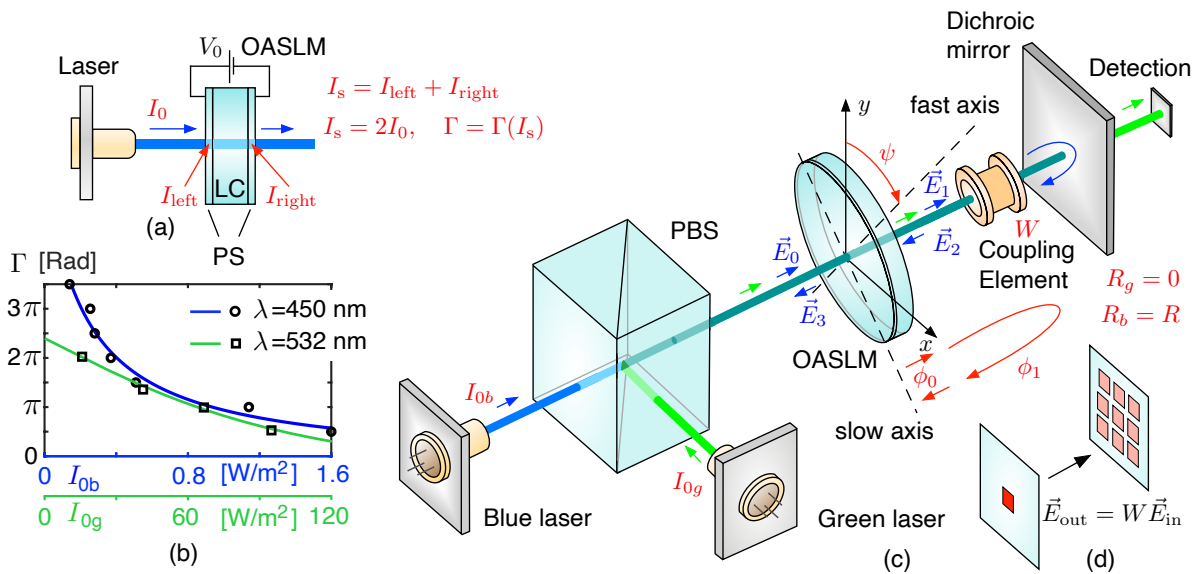} 
	\caption{(a) OASLM illuminated by a light source; (b) Experimentally obtained retardation for blue ($\lambda_{b}=450$~nm, black circles) and green ($\lambda_{g}=532$~nm, black squares) light, measured at $V_0=9~$V.
		Fitting with $\Gamma(I_0)=(2\alpha I_0+\beta)^{-1}+\gamma$ results in $\alpha_{b} = 0.117$, $\beta_{b} = 0.052$, $\gamma_{b} = -0.55$ (blue solid line) and $\alpha_{g} = 98.5\times 10^{-4}$, $\beta_{g} = 0.0486$, $\gamma_{g} = -13.17$ (green solid line).
		(c) Schematic of the two-color OSALM experiment using a single PS-layer on the left side.
		Blue light is reflected by a dichroic mirror and creates feedback, green light leaves the system and can be used for state detection.
		(d) Illustration of the coupling element action for the case of local interaction.}
	\label{fig1}
\end{figure*}  

However, the physical composition and construction principles of reflective SLMs electrically (EASLMs) or optically (OASLM) addressing result in constrains.
Reflective illumination makes coherent interference between the optical state variable and coupling fields challenging, yet it substantially enriches the range of dynamical behaviour accessible to the system.
Such interference is straightforward with transmissive OASLMs.
Furthermore, EASLMs require extensive control hardware, which implies additional energy consumption that is a disadvantage for high energy efficient computing.
Transmissive OASLMs are therefore a smart and powerful solution for a variety of fundamental and technological challenges. 

In this paper we show that a variety of dynamical systems can be implemented using transmissive OASLMs under two color illumination coupled to an external cavity to realize coupling.
We identify the corresponding bifurcation scenarios by deriving normal forms through the Taylor-expansion of the equations governing the system dynamics.
These we then associate to the conditions specifically realizing pitchfork, transcritical and saddle-node bifurcations of steady states, however, intermediate configurations are possible.
Simply by adjusting the relative intensities of the two-colored illumination intensities to continuously transitioning between these different scenarios.
Here, one color encodes the system's dynamical state and its optical coupling, while the other color realizes a constant DC control parameter.
Our OASLM model is inspired by I. Abdulhalim et al. \cite{kirzhner2014}, and we used one of their proof of concept devices for obtaining the device parameters used in analytical derivations and numerical modeling.

\section{Experimental characterization and model development}

The OASLM we used for our basic device parameter characterization leverages a nematic liquid crystal (LC) layer located between two a-As$_2$S$_3$ chalcogenide thin films (60 nm) that simultaneously function as photosensitive (PS) and alignment layers [Fig.~\ref{fig1}~(a)].
Crucially, the PS layer absorption and hence photoresponse depends on the incident light wavelength.
Amorphous As$_2$S$_3$ films are almost transparent and hence insensitive for red light.
However, even for our thin PS layers, strong absorption of blue light modifies the OASLM's state for illumination intensities as low as 10 nW/mm$^2$.
The nematic LC (Merck E44) exhibits ordinary and extraordinary refractive index $n_o$ and $n_e$, respectively, characterized by birefringence $\Delta n = n_e-n_o$.
Propagation through the LC layer results in optical phase retardation $\Gamma = \frac{2\pi d}{\lambda} \Delta n$, where $\lambda$ is the incident light wavelength, $d$ is the LC thickness.
For our basic characterization we connect the OASLM to a DC-power source resulting in a voltage drop across the LC layer that is uniform without illumination.
However, illumination [Fig.~\ref{fig1}~(a)] spatially modifies the PS's conductivity and local voltage across the LC layer.
Due to their induced dipole moment, LC molecules change their orientation in response that results in a spatial birefringence distribution that is a function of the optical illumination profile.

Our device has two PS layers, and hence the relevant optical intensity inducing retardation $\Gamma$ is $I_\text{s}=I_{\text{left}}+I_{\text{right}}$, see Fig.~\ref{fig1}~(a), which when ignoring the small optical absorption of the PS layers becomes $I_\text{s}=2I_0$.
The experimentally obtained retardation $\Gamma$ on the incident light intensity $I_{0}$ is depicted in Fig.~\ref{fig1}~(b) for blue (SHD4580MG, $\lambda_{b}=450$ nm) and green (DJ532-10, $\lambda_{g}=532$ nm) laser illumination.
The OASLM is substantially more sensitive to blue than green illumination, and we fit the dependency with $\Gamma(I_0)=(2\alpha I_0+\beta)^{-1}+\gamma$ (solid lines in Fig.~\ref{fig1}~(b)) \cite{kirzhner2014}.
In our study we found that, besides doubling the responsivity of the device, the second PS-layer does not increase the dynamical complexity.
For the sake of simplifying the dynamical equations, in the rest of the manuscript we therefore only consider an OASLM with a single PS-layer located on the left side of the LC-layer.
In our generic setup, depicted in Fig.~\ref{fig1}~(c), the PS-layer is simultaneously illuminated from the left direction by a blue and a green laser.
After traversing the OASLM, a dichroic mirror transmits the green light beam but reflects the blue with reflectivity $R$, which then interferes with the original blue illumination.
Generally, optical interference, i.e. a temporal beating originating from the superposition blue and green light, can be ignored.

Here, we consider horizontally polarised blue illumination, which we express as $\vec{E}_0~=~[E_{0}, 0]$, and the OASLM's rotation relative to this polarization is $\psi=\pi n/2$.
Using Jones matrix calculus, the field arriving after the second transition through the OASLM is $\vec{E}_3~=~R\exp(i (2\phi_0+\phi_1+2\Gamma))[E_0, 0]$, where $\phi_0$ is the constant retardation induced by the OASLM without illumination and $\phi_1$ the retardation accumulated in the external cavity round-trip, $i$ is the imaginary unit.
Then the resulting blue light field at the left PS-layer is $\vec{E}_{\text{b}}=\vec{E}_0+\vec{E}_3$ and the blue light intensity is $I_{\text{b}}=| \vec{E}_{\text{b}}|^2$.
Finally, optical feedback is considered instantaneous relative to the OASLM's response time $\varepsilon$.
Based on this, and for the case of homogeneous illumination and feedback, the equation for the temporal evolution of the blue light retardation becomes:
\begin{equation}
	\label{single_oscillator}
	\begin{array}{l}
		\varepsilon \dfrac{d\Gamma}{dt}=-\Gamma+\dfrac{1}{\alpha_{\text{b}} I_{\text{b}}(\Gamma)+\tilde{\alpha}_g I_{0\text{g}}+\beta}+\gamma,\\
		I_{\text{b}}(\Gamma)=I_{0\text{b}}\Big\{1+R^2+ 2 R\cos(2\phi_0+\phi_1+2\Gamma)       \Big\},
	\end{array}
\end{equation}
and $\tilde{\alpha}_g = \frac{\lambda_{\text{g}}}{\lambda_{\text{b}}}\alpha_g$ is the retardation effect of $I_{0\text{g}}$ on the blue signal.
In the rest of the manuscript we use fixed parameters $R=0.95$, $\alpha_{\text{b}}=0.117$, $\alpha_{\text{g}}=98.5\times 10^{-4}$, $\beta=0.052$, $\gamma=-0.55$, $\varepsilon=1$, and varying $I_{0\text{b}}$ and $I_{0\text{g}}$.
These correspond to typical values of experiments or were obtained through fitting the retardation of our OASLM at different wavelengths.
In what follows we generally set $2\phi_0+\phi_1=2\pi n$ ($n\in \mathbb{Z}$).
This significantly simplifies Eq. (\ref{taylor_series}) and the system is still able to obtain the pitchfork, saddle-node and transcritical normal forms.
Mathematical derivations presented in the next section are combined with numerical simulation of Eq. (\ref{single_oscillator}) using the Heun method \cite{mannella2002} with the time step $\Delta t=10^{-3}$.

\section{Taylor series and bifurcations}

Equation (\ref{single_oscillator}) is a one-dimensional oscillator expressed in real-valued variables.
Three kinds of bifurcation transition can be observed in such class of dynamical systems: the pitchfork bifurcation, the saddle-node bifurcation and the transcritical bifurcation with their corresponding normal forms of $dx/dt=bx+dx^3$ (pitchfork bifurcation), $dx/dt=a+cx^2$ (saddle-node bifurcation) and $dx/dt=bx+cx^2$ (transcritical bifurcation).
In order to express our system in a relatable form, we approximate the right-hand side of Eq. (\ref{single_oscillator}) as a cubic function $f_{\text{T}}(\Gamma)$ via Taylor series
\begin{equation}\label{eq:Taylor}
	f_{\text{T}}(\Gamma)=a+b(\Gamma-\Gamma_0)+c(\Gamma-\Gamma_0)^2+d(\Gamma-\Gamma_0)^3    
\end{equation}
with $a=f(\Gamma_0)$, $b=f'(\Gamma_0)$, $c=\frac{1}{2}f''(\Gamma_0)$, $d=\frac{1}{6}f'''(\Gamma_0)$ and $f'=\frac{df}{d\Gamma}$, $f''=\frac{d^2f}{d\Gamma^2}$ and $f'''=\frac{d^3f}{d\Gamma^3}$.
The corresponding factors associated to the relevant polynomial orders are
\begin{equation}
	\label{taylor_series}
	\begin{array}{l}
		f_{\text{T}}(\Gamma)=a+b(\Gamma-\Gamma_0)+c(\Gamma-\Gamma_0)^2+d(\Gamma-\Gamma_0)^3, \\
		a=-\Gamma_0+\left(\alpha_{\text{b}} I_{\text{b}}+\tilde{\alpha}_{\text{g}} I_{\text{0g}}+\beta \right)^{-1}+\gamma,\\
		b=-1+\xi I_{0\text{b}}\sin(2\Gamma_0)\left(\alpha_{\text{b}} I_{\text{b}}+\tilde{\alpha}_{\text{g}} I_{\text{0g}}+\beta \right)^{-2},\\
		c=\xi I_{0\text{b}}\cos(2\Gamma_0)\left(\alpha_{\text{b}} I_{\text{b}}+\tilde{\alpha}_{\text{g}} I_{\text{0g}}+\beta \right)^{-2}\\
		+(\xi I_{0\text{b}})^2 \sin^2(2\Gamma_0)\left(\alpha_{\text{b}} I_{\text{b}}+\tilde{\alpha}_{\text{g}} I_{0\text{g}}+\beta \right)^{-3},\\
		d=-\dfrac{8}{3}\xi I_{0\text{b}}\sin(2\Gamma_0)\left(\alpha_{\text{b}} I_{\text{b}}+\tilde{\alpha}_{\text{g}} I_{0\text{g}}+\beta \right)^{-2}\\
		+(\xi I_{0\text{b}})^2\sin(4\Gamma_0)\left(\alpha_{\text{b}} I_{\text{b}}+\tilde{\alpha}_{\text{g}} I_{0\text{g}}+\beta \right)^{-3}\\
		+(\xi I_{0\text{b}})^3 \sin^3(2\Gamma_0)\left(\alpha_{\text{b}} I_{\text{b}}+\tilde{\alpha}_{\text{g}} I_{0\text{g}}+\beta \right)^{-4},
	\end{array}
\end{equation}
where $\xi= 4\alpha_{\text{b}} R$ is an optical responsivity scaling factor of optical feedback.
Crucially, we control $\Gamma_0$ through $I_{0\text{b}}$ and $I_{0\text{g}}$, which is the mechanism we use for tuning Taylor series coefficients to zero in order to transform $f_{\text{T}}$ into a desired normal form.

\subsection{Pitchfork bifurcation}

\begin{figure}[t]
	\centering
	\includegraphics[width=0.48\textwidth]{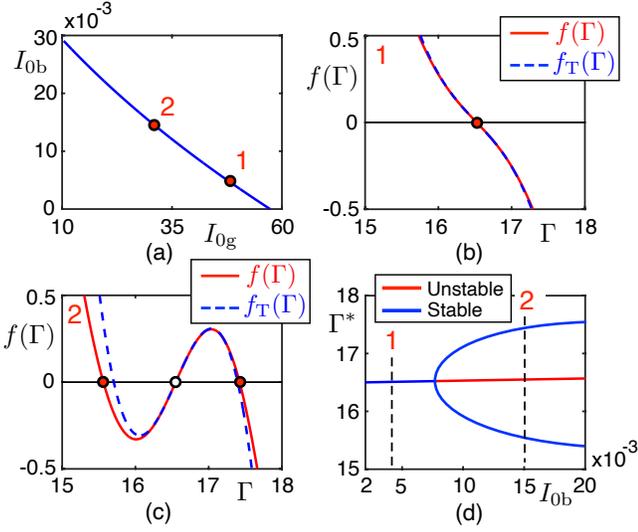} 
	\caption{Pitchfork bifurcation in system (\ref{single_oscillator}) when $I_{0\text{b}}$ and $I_{0\text{g}}$ vary as in panel (a): Panels (b) and (c) display the evolution of the right-hand side function $f(\Gamma)$ of Eq.~\ref{single_oscillator} and its Taylor series approximation $f_{\text{T}}(\Gamma)$ on the example of points 1 ($\Gamma_0=16.5$, $I_{0\text{b}}=0.0043$, $I_{0\text{g}}=48.7$) and 2 ($\Gamma_0=16.6$, $I_{0\text{b}}=0.0151$, $I_{0\text{g}}=30.1$) while panel (d) illustrates the full phase-parametric diagram (here $\Gamma^*$ denotes the steady state coordinates). Parameters are: $R=0.95$, $\alpha_{\text{b}}=0.117$, $\alpha_{\text{g}}=98.5\times 10^{-4}$, $\beta=0.052$, $\gamma=-0.55$, $2\phi_0+\phi_1=2\pi$, $\varepsilon=1$.}
	\label{fig2}
\end{figure}  

Function $f_{\text{T}}(\Gamma)$ coincides with the pitchfork bifurcation normal form when the Taylor series coefficients $a$ and $c$ (see Exps. (\ref{taylor_series})) are zero.
The condition $a=0$ gives rise to:
\begin{equation}
	\label{pitchfork_condition_a=0}
	\left(\alpha_{\text{b}} I_{\text{b}}+\tilde{\alpha}_{\text{g}} I_{\text{0g}}+\beta \right)^{-1}=\Gamma_0 - \gamma. \\
\end{equation}
Substituting (\ref{pitchfork_condition_a=0}) into the second condition $c=0$, one obtains:
\begin{equation}
	\label{pitchfork_condition_c=0}
	\begin{array}{l}
		\xi I_{0\text{b}}\cos(2\Gamma_0)(\Gamma_0-\gamma)^{2}\\
		+(\xi I_{0\text{b}})^2 \sin^2(2\Gamma_0)(\Gamma_0-\gamma)^{3}=0,
	\end{array}
\end{equation}
which allows to express the incident blue light intensity parameter as a function of $\Gamma_0$:
\begin{equation}
	\label{I_0b_pitchfork}
	I_{0\text{b}}(\Gamma_0)=-\dfrac{\cos(2\Gamma_0)}{\xi\sin^2(2\Gamma_0)(\Gamma_0-\gamma)}.
\end{equation}
Substituting (\ref{I_0b_pitchfork}) back into (\ref{pitchfork_condition_a=0}) allows to similarly express $I_{0\text{g}}$ as a function of $\Gamma_0$:
\begin{equation}
	\label{I_0g_pitchfork}
	\begin{array}{l}
		\tilde{\alpha}_{\text{g}} I_{0\text{g}}(\Gamma_0)= \dfrac{1}{\Gamma_0-\gamma}
		-\beta
		\\
		-\alpha_{\text{b}}I_{0b} \left\{1+R^2+2R\cos(2\Gamma_0)    \right\}.
	\end{array}
\end{equation}
$I_{0\text{g}}$ therefore becomes a function of $I_{0\text{b}}$ which in turn is determined through Eq. (\ref{I_0b_pitchfork}) as a function of $\Gamma_0$.
Figure~\ref{fig2}~(a) shows the solution of Eqs.~(\ref{I_0b_pitchfork}, \ref{I_0g_pitchfork}), and the resulting solutions at characteristic points 1 and 2 for $f_{\text{T}}(\Gamma)$ and $f(\Gamma)$ are shown in Fig.~ \ref{fig2}~(b) and (c), respectively.

Transitioning from point 1 to point 2 in the $(I_{0b},I_{0g})$ the system bifurcates from a single stable fixed point (c.f. Fig.~\ref{fig2}~(b)) to a pair of stable fixed point separated by an unstable fixed point (c.f. Fig.~\ref{fig2}~(c)).
For a pitchfork bifurcation, this transition is according to a cubic function with $\Gamma_0$ as its central point.
Setting $a=c=0$ ensured that Taylor approximation $f_{\text{T}}(\Gamma)$ is strictly of cubic order (blue data in Fig.~\ref{fig2}~(b,c)), and comparison to the system's full analytical description $f(\Gamma)$ (red data in Fig.~\ref{fig2}~(b,c)) shows that this approximation is very accurate.
In Fig.~\ref{fig2}~(d) we show the system's phase-parametric bifurcation scenario by tracking the stable (blue data) and unstable (red data) fixed points of the non-approximated $f(\Gamma)$.
Beyond the critical value of $I_{o\text{b}}$, the system shows a parabolically increasing distance between two stable fixed points, which are separated by the unstable fixed point.
This hence confirms that the OASLM with coherent feedback and for an adjustment of $I_{0g}$ relative to $I_{0b}$ according to Eq.~(\ref{I_0g_pitchfork}) causes the system evolve according to a pitchfork bifurcation.

\subsection{Saddle-node bifurcation}

\begin{figure}[t]
	\centering
	\includegraphics[width=0.48\textwidth]{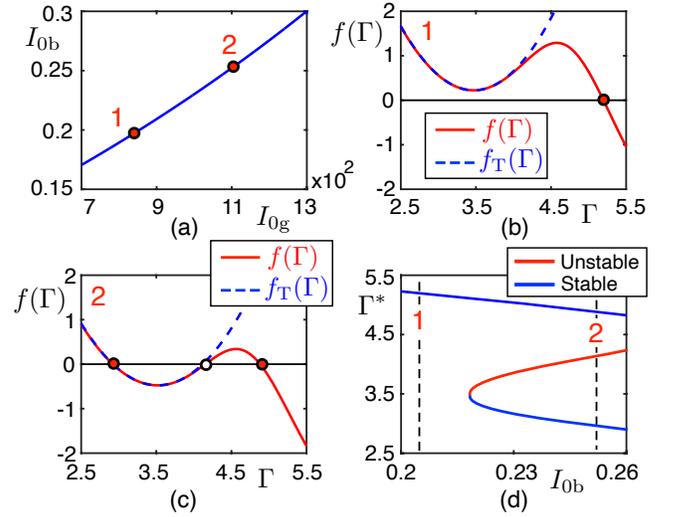} 
	\caption{Saddle-node bifurcation in system (\ref{single_oscillator}) when $I_{0\text{b}}$ and $I_{0\text{g}}$ vary as in panel (a): Panels (b) and (c) display the evolution of the right-hand side function $f(\Gamma)$ and its Taylor series approximation $f_{\text{T}}(\Gamma)$ on the example of points 1 ($\Gamma_0=3.47$, $I_{0\text{b}}=0.205$, $I_{0\text{g}}=877$) and 2 ($\Gamma_0=3.52$, $I_{0\text{b}}=0.254$, $I_{0\text{g}}=1106$) while panel (d) illustrates the full phase-parametric diagram (here $\Gamma^*$ denotes the steady state coordinates). Other parameter values are the same as in Fig. \ref{fig2}.}
	\label{fig3}
\end{figure}  

Following the same approach we derive the conditions for the saddle-node bifurcation, for which we require $b=d=0$.
The first condition leads to
\begin{equation}
	\label{saddle_node_condition_b=0}
	\begin{array}{l}
		\xi I_{0\text{b}} \sin(2\Gamma_0) \left(\alpha_{\text{b}} I_{\text{b}}+\tilde{\alpha}_{\text{g}} I_{\text{0g}}+\beta \right)^{-2}=1,
	\end{array}
\end{equation}
which is inserted into the condition $d=0$.
After that $d=0$ can be represented as a quadratic equation for the variable $k=\left(\alpha_{\text{b}} I_{\text{b}}+\tilde{\alpha}_{\text{g}} I_{\text{0g}}+\beta \right)$:
\begin{equation}
	\label{saddle_node_condition_d=0}
	\begin{array}{l}
		-\dfrac{2}{3}+2\cot(2\Gamma_0)k+k^2=0.
	\end{array}
\end{equation}
Taking into account that light intensities $I_{0\text{b}}$ and $I_{0\text{g}}$ cannot be negative, $k$ possesses only positive values.
Solving Eq. (\ref{saddle_node_condition_d=0}), one obtains the following equality:
\begin{equation}
	\label{saddle_node_condition_d=0_solution}
	\begin{array}{l}
		\left(\alpha_{\text{b}} I_{\text{b}}+\tilde{\alpha}_{\text{g}} I_{\text{0g}}+\beta \right)=\\
		-\cot(2\Gamma_0)+\sqrt{\cot^2(2\Gamma_0)+\frac{2}{3}}.
	\end{array}
\end{equation}
After substituting Eq.~(\ref{saddle_node_condition_d=0_solution}) into Eq.~(\ref{saddle_node_condition_b=0}), parameter $I_{0\text{b}}$ can finally be expressed as a function of $\Gamma_0$:
\begin{equation}
	\label{I_0b_saddle_node}
	I_{0\text{b}}(\Gamma_0)=\dfrac{\left(\sqrt{\cot^2(2\Gamma_0)+\dfrac{2}{3}} -\cot(2\Gamma_0)\right)^2}{\xi \sin(2\Gamma_0)}.
\end{equation}
We continue by substituting Eq.~(\ref{I_0b_saddle_node}) into (\ref{saddle_node_condition_b=0}), which allows solving the conditions for $I_{0\text{g}}$:
\begin{equation}
	\label{I_0g_saddle_node}
	\begin{array}{l}
		\tilde{\alpha}_{\text{g}} I_{0\text{g}}(\Gamma_0)=\sqrt{\cot^2(2\Gamma_0)+\dfrac{2}{3}}-\beta\\
		-\alpha_{\text{b}}I_{0b} \left\{1+R^2+2R\cos(2\Gamma_0) \right\} -\cot(2\Gamma_0).
	\end{array}
\end{equation}

After having derived $I_{0g}$ as a function of $\Gamma_{0}$ and $I_{0b}$ according to Eq.~(\ref{I_0b_saddle_node}) and (\ref{I_0g_saddle_node}), we follow the same logic as in our analysis of the pitchfork bifurcation.
The relationship between $I_{0\text{b}}(\Gamma_0)$ and $I_{0\text{b}}(\Gamma_0)$ is illustrated in Fig.~\ref{fig3}~(a).
We again show the system before and after bifurcation in Fig.~\ref{fig3}~(b) and Fig.~\ref{fig3}~(c), respectively, and the comparison between the quadratic $f_{\text{T}}(\Gamma)$ (blue data) and the full description $f(\Gamma)$ (red data) shows that in the vicinity of the relevant fixed point the Taylor series excellently approximates the system's behaviour.
Finally, Fig.~\ref{fig3}~(d) shows the system's full bifurcation diagram according to $f(\Gamma)$, which perfectly reproduces the classical behaviour of a saddle-node bifurcation.

\subsection{Transcritical bifurcation}

The conditions for a transcritical bifurcation are $a=d=0$.
The first condition results in Eq.~(\ref{pitchfork_condition_a=0}) previously derived for the pitchfork bifurcation, which is substituted into the condition $d=0$, which results in a cubic equation for variable $k=\xi I_{0\text{b}} \sin(2\Gamma_0) (\Gamma_0-\gamma)$:
\begin{equation}
	\label{transcritical_condition_d=0}
	\begin{array}{l}
		k\left(k^2+2\cot(2\Gamma_0)k-\dfrac{2}{3}\right)=0,
	\end{array}
\end{equation}
which has three roots.
A transcritical bifurcation occurs at positive values $I_{0\text{b}}$ and $I_{0\text{g}}$ only for the solution $k=-\cot(2\Gamma_0)+ \sqrt{\cot^2(2\Gamma_0)+\frac{2}{3}}$, and $I_{0\text{b}}$ can be expressed and represented as a function of $\Gamma_0$:
\begin{equation}
	\label{I_0b_transcritical}
	\begin{array}{l}
		I_{0\text{b}}(\Gamma_0)=\dfrac{-\cot(2\Gamma_0)+\sqrt{\cot^2(2\Gamma_0)+\dfrac{2}{3}}}{\xi \sin(2\Gamma_0)(\Gamma_0-\gamma)}.
	\end{array}
\end{equation}
As before, we substitute Eq.~(\ref{I_0b_transcritical}) into Eq.~(\ref{pitchfork_condition_a=0}) to finally obtain
\begin{equation}
	\label{I_0g_transcritical}
	\begin{array}{l}
		\tilde{\alpha}_{\text{g}}I_{0\text{g}}(\Gamma_0)= \left(\dfrac{1}{\Gamma_0-\gamma}-\beta\right)\\
		-\alpha_b I_{0b} \Big\{1+R^2+2R\cos(2\Gamma_0) \Big\}.
	\end{array}
\end{equation}

\begin{figure}[t]
	\centering
	\includegraphics[width=0.48\textwidth]{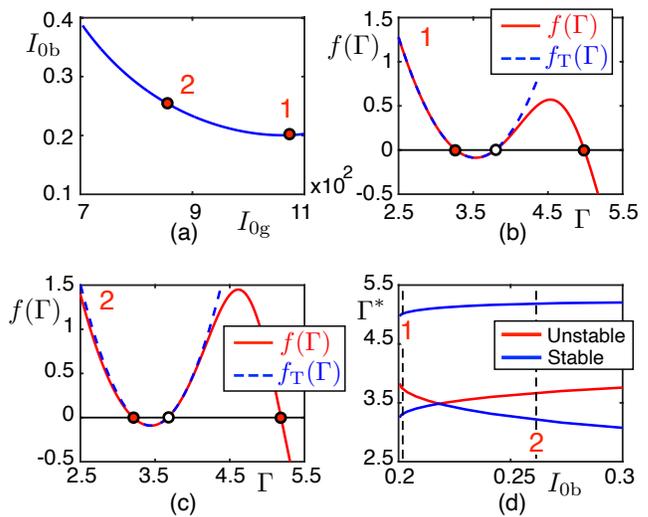} 
	\caption{Transcritical bifurcation in system (\ref{single_oscillator}) when $I_{0\text{b}}$ and $I_{0\text{g}}$ vary as in panel (a): Panels (b) and (c) display the evolution of the right-hand side function $f(\Gamma)$ and its Taylor series approximation $f_{\text{T}}(\Gamma)$ on the example of points 1 ($\Gamma_0=3.2$, $I_{0\text{b}}=0.20084$, $I_{0\text{g}}=1079$) and 2 ($\Gamma_0=3.4845$, $I_{0\text{b}}=0.21803$, $I_{0\text{g}}=943$) while panel (d) illustrates the full phase-parametric diagram (here $\Gamma^*$ denotes the steady state coordinates). Other parameter values are the same as in Figs. \ref{fig2} and \ref{fig3}.}
	\label{fig4}
\end{figure}

Our following analysis is identical to the two cases shown before.
Figure~\ref{fig4}~(a) depicts the relationship between $I_{0\text{b}}$ and $I_{0\text{g}}$, Fig.~\ref{fig4}~(b,c) shows $f(\Gamma)$ and $f_T(\Gamma)$ before and after bifurcation. 
Finally, Fig.~\ref{fig4}~(d) confirms the classical transcritical bifurcation scenario of the system.

\section{Generic nonlinear oscillator networks}

In the previous sections we have demonstrated that a OASLM subjected to coherent feedback is capable to realize all bifurcations possible with a real-valued nonlinear oscillator.
We now change the general context.
Now, horizontally polarised incident light is considered as a set of $n$ pixels, and we express its optical field at different positions in the form of the two $n$-dimensional vectors
\begin{equation}
	\label{network_model_incident_light}    
	\vec{E}_{0x}~=~
	\begin{bmatrix}
		E_0\\
		\vdots\\
		E_0
	\end{bmatrix},\quad\quad
	\vec{E}_{0y}~=~
	\begin{bmatrix}
		0\\
		\vdots\\
		0
	\end{bmatrix}.
\end{equation}
Between OASLM and the dichroic mirror we now place optical components for coupling such pixels, e.g. a diffractive element.
The optical feedback field $\vec{E}_{3}$ at the $k$-th position becomes
\begin{equation}
	\label{network_model_returned_light}  
	\vec{E}^{3x}_{k}~=~E_0 R\sum_{l=1}^{n} W_{k,l}\exp(2i\Gamma_l)
\end{equation}
where $\mathbf{W}$ is a $n\times n$ coupling matrix without birefringence (i.e. $\vec{E}^{3y}_{k}=0$) and $\vec{\Gamma}$ is the $n$-dimensional spatial birefringence distribution of the OASLM.
The optical intensity on the PS in such a optical feedback-coupled network takes the form
\begin{equation}
	\label{network_model}  
	\begin{array}{l}
		I^{\text{b}}_k=I_{0\text{b}} \Big\{ 1+2 R \sum\limits_{l=1}^{n} W_{k,l} \cos(2\vec{\Gamma_{l}}) \\
		+R^2\Big(\sum\limits_{l=1}^n W_{k,l}^2 +2\sum\limits_{\substack{l,m=1,\\l\neq m}}^n W_{k,l}W_{k,m}\cos(2(\Gamma_l-\Gamma_m))\Big)\Big\}.
	\end{array}
\end{equation}

Optical coupling matrices can realize a variety of coupling topologies.
For instance, regular matrices can be created using diffractive optical elements \cite{brunner2015,bueno2018}, while scattering media results in random matrices\cite{rafayelyan2020}. 
We consider local coupling by splitting any pixel's field into 9 identical beams by including a diffractive beam splitter in the external cavity \cite{brunner2015}, see Fig.~\ref{fig1}~(d).
Furthermore, we investigate random coupling with different degrees of connectivity.

\begin{figure}[t]
	\centering
	\includegraphics[width=0.48\textwidth]{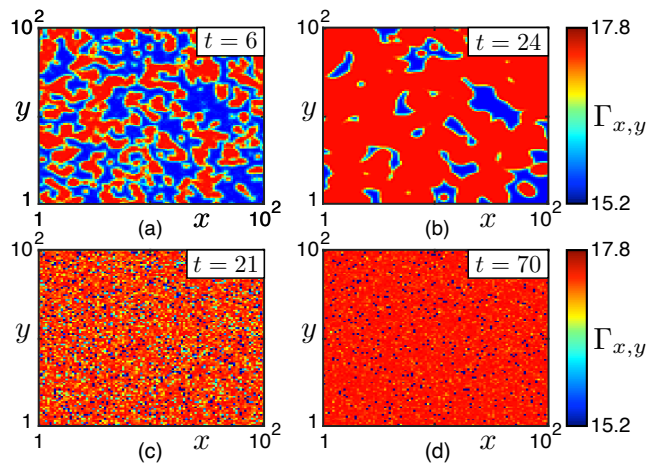} 
	\caption{Spatial evolution of system (\ref{network_model}) starting from random initial conditions for local coupling (panels (a)-(b)) and for sparse random  matrix $W$ (panels (c)-(d)). Light intensity parameters are $I_{0\text{b}}=0.01506$, $I_{0\text{g}}=30.1$. Other parameter values are the same as in Figs. \ref{fig2}-\ref{fig4}.}
	\label{fig5}
\end{figure}

First, we model Eq.~(\ref{network_model}) with local coupling and in the bistable regime with $I_{0\text{b}}$ and $I_{0\text{g}}$ as in point 2 in Fig. \ref{fig2} (c)).
Numerical simulations start from random initial conditions for $\Gamma$, and the system evolves into spatial domains comprising of either steady state $\Gamma^*_1$ or $\Gamma^*_2$ (blue and red domains in Fig.~\ref{fig5}~(a)).
The temporal evolution exhibits coarsening: either $\Gamma^*_1$ or $\Gamma^*_2$ invades the entire available space, see Fig. \ref{fig5}~(b).
Any deviation of $I_{0\text{b}}$ and $I_{0\text{g}}$ from the curve in Fig.~\ref{fig2}~(a) induces asymmetry, which we find to accelerate coarsening and to break its symmetry, which hence is in agreement with the corresponding literature \cite{engel1985,loeber2014}.

We then study the same system with random coupling with real and positive entries that are uniformly distributed random numbers normalized such that column-sums of $\mathbf{W}$ are unity.
If coupling is global, the network transitions close to either $\Gamma_1^{*}$ or $\Gamma_2^{*}$.
However, if $W$ is highly sparse, the system-attractor corresponds to the non-uniform spatial state as shown in Fig.~\ref{fig5}~(c,d) for $W$ containing 99.9$\%$ of zero elements.

In contrast to the pitchfork bifurcation, saddle-node bifurcations do not allow controlling the system's symmetry.
However, one can move $f(\Gamma)$ up and down (as in Fig.~\ref{fig3}), and thereby transition from monostability to bistability, and to contract or extend the basins of attraction of the coexisting steady states after the bifurcation.
Similarly, transcritical bifurcations allows swapping stable for unstable fixed-points, and vice-versa: the stability of steady states changes at the transcritical bifurcation moment.

\section{Conclusions}

We developed the general nonlinear-dynamical model for transmissive OASLMs with coherent optical feedback that allows to implement large-scale networks via optical connections. 
We have demonstrated that the single-oscillator model can exhibit the pitchfork, transcritical and saddle-node bifurcations of steady states, and our concept is highly versatile as the continuous tuning between these bifurcation scenarios only required adjustment of the intensity ratio of our two-color illumination.
Here we used the example of very different wavelength for both colors, which results in two orders of magnitude intensity-differences due to the changing PS sensitivity.
This is easily achievable with attenuators, however, more comparable values are straightforward by using two color illumination with closer wavelengths, which only requires changing the dichroic mirror.

We created network models by deriving the dynamical system for optical coupling matrices.
For local coupling in the case of oscillators with the pitchfork bifurcation, we find that domains of bistability expand, which corresponds to the effect of coarsening.
In the presence of random coupling the system tends to nonuniform spatial states.
Similar situations are observed for the case of saddle-node and transcritical bifurcations, where however we can control the ratio and stability of domains associated with either steady state.
Networks of bistable elements exhibiting the pitchfork bifurcations are of relevance in artificial intellingence \cite{stern2014,vecoven2021}.
Similarly, networks of elements exhibiting saddle-node bifurcations include neurons neuron-models \cite{roxin2016,tartaglia2017,arthur2011}, while transcritical systems present interacting subnetworks of spiking oscillators \cite{lagzi2019}.

Based on experimental characterizations of OASLMs, we find such a system is capable to implement up to $10^4$ nodes per mm$^2$ only requiring illumination intensities as low as 10 nW/mm$^2$ at $\lambda_{\text{b}}=450~$nm.
Consequently, such OASLMs are promising candidates for implementing autonomous novel next generation optical computing architectures \cite{bueno2018,pierangeli2019} with ultra low energy consumption.

\section*{DATA AVAILABILITY}
The data that support the findings of this study are available from the corresponding author upon reasonable request.

\section*{Acknowledgements}
This work has been supported by the EIPHI Graduate School (contract ANR-17-EURE-0002) and the Bourgogne-Franche-Comt\'e Region and H2020 Marie Skłodowska-Curie Actions (MULTIPLY, NO 713694). 

\bibliography{aipsamp}

\end{document}